\documentclass{aa}  

\usepackage[colorlinks=true]{hyperref}
\hypersetup{linkcolor=blue, citecolor=blue}
\usepackage[draft]{todonotes}
\usepackage{amsmath}

\usepackage{graphicx}
\usepackage{txfonts}
\begin{document} 
%\newenvironment{alignteo}%
  %{\empheq[box=\tcbhighmath]{align}}
  %{\endempheq}
%
% Units
%
\newcommand{\ColDens}{cm$^{-2}$}
\newcommand{\kms}{km~s$^{-1}$}
\newcommand{\mum}{$\mu$m}
%
% Polarization parameters
%
\newcommand{\lmax}{$\lambda_{max}$}
\newcommand{\pmax}{$p_{max}$}
\newcommand{\PolPlanck}{$p_{353}$}
%
% Extinction
%
\newcommand{\Rv}{$R_{V}$}
\newcommand{\EBV}{$E (B - V)$}
\newcommand{\Av}{$A_{V}$}
%
% Grain properties
%
\newcommand{\amin}{$\alpha_{-}$}
\newcommand{\amax}{$\alpha_{+}$}
\newcommand{\EVPAvariability}{$\Delta \theta$}
%red color
\newcommand{\rs}[1]{\textcolor{red}{#1}}

   \title{Challenges in constraining dust properties from starlight polarization}

   \subtitle{}

   \author{Raphael Skalidis}

    \institute{ Owens Valley Radio Observatory, California Institute of Technology, MC 249-17, Pasadena, CA 91125, USA\thanks{Hubble fellow}
    			\and 
	Faculty of Engineering and Natural Sciences, Sabancı University, Orhanlı-Tuzla 34956, Istanbul, Turkey 
	\\ \email{rafail.skalidis@sabanciuniv.edu} }

   \date{Received; accepted }

% \abstract{}{}{}{}{} 
% 5 {} token are mandatory
 
  \abstract
   {Dust polarization, which comes from the alignment of aspherical grains to magnetic fields, has been widely employed to study the interstellar medium (ISM) dust properties. The wavelength dependence of the degree of optical polarization, known as the Serkowski relation, was a key observational discovery that advanced grain modeling, and alignment theories. However, it was recently shown that line-of-sight (LOS) variations in the structure of the ISM or the magnetic field morphology contaminate the constraints extracted from fits to the Serkowski relation. These cases can be identified by the wavelength-dependent variability in the polarization angles. 
   }
   {We aim to assess how reliably the Serkowski parameters and polarization efficiency can be constrained from starlight polarization, considering the effects of limited wavelength coverage, prior assumptions, and LOS variations in magnetic field morphology.}
   {We combine multi-wavelength starlight polarization measurements from a compiled catalogue with Gaia stellar parallaxes and three-dimensional dust extinction maps. We fit the Serkowski relation using both uniform priors and informed priors that favor the Wilking relation, and compare the resulting parameter constraints.} 
   {We find that width of the polarization spectrum ($K$) is particularly sensitive to the adopted prior, whereas the uncertainties in \pmax\ and \lmax\ are comparable for the informed and uniform priors. The informed-prior fits favor the reference Wilking relation. In contrast, the uniform-prior fits yield a broader distribution of $K$: while some measurements remain consistent the Wilking relation, several exhibit abnormally narrow polarization profiles where $K \gtrsim 3$. These high-$K$ candidates span a wide range of polarization efficiencies, indicating that narrow profiles are not systematically driven by enhanced grain alignment. Beyond explanations based on grain properties, LOS superposition of multiple clouds can reproduce such narrow profiles even when individual component follows the Wilking relation.}
   {Sparse wavelength sampling, assumptions about spectral shape, and the three-dimensional structure of the ISM limit the interpretation of fitted polarization parameters in terms of intrinsic grain properties. Broad-wavelength spectropolarimetry, combined with stellar distances, three-dimensional dust maps, and distance-resolved magnetic field structure, can improve constraints on polarization spectra and help distinguish grain-property variations from integration effects.}


   \maketitle

%
%-------------------------------------------------------------------
\section{Introduction}

Despite accounting for only $\sim 1\%$ of the total (interstellar medium) ISM mass, dust is omnipresent and integral to numerous processes, such as protostar formation \citep{mouschovias_1996.dust.protostar.formation}, molecular cloud dynamics \citep{hopkins_2014.dgr.variations.formation.metal.rich.stars,squire_hopkins_2018.drag.instablities,hopkins_2022.dust.in.the.wind,hennebelle_2023.dust.inertia.alfven.wave.propagation}, cloud thermodynamics, astrochemistry \citep{goldsmith_2001.cooling.cores,bialy_sternberg_2019.thermal.phases.hi.clouds}, and Galaxy evolution. It is thus imperative to understand the ISM dust grain properties

Dust polarization is one of the most widely used and accurate methods for investigating the ISM dust grain properties. This effect comes from the interaction between aspherical dust grains and magnetic fields  \citep{andersson_2015.review}. Dust polarization provides important constraints on the ISM grain properties \citep{guillet_2018.dust_models.Planck.compatibility,hensley_draine2021.dust.grains.constraints,hensley_draine_2023.pah.astrodust} and magnetic fields \citep{davis_1951,chandra_fermi,ostriker_2001,hildebrand_2009,houde_2009,cho_2016.dcf.modifation,chen_2022.dcf.modified, panopoulou_2015, panopoulou_2016,versteeg_2023.polarization.map,panopoulou_2019_tom,hu_lazarian_2023.Bfield.inclination.angle.polarization, pelgrims_2023.tom.singleLOS, schmaltz.hu.laz_2024.3Dtomography.VelocityGradients.Polarization, doi_2024.3D.magnetic.field.sagitarius.arm, skalidis_pelgrims_2019, skalidis_tassis_2021,skalidis_2021_Bpos}. Single-band polarization measurements suffice for magnetic field studies \citep[e.g.,][]{skalidis_2022,skalidis_2023.polaris.flare}, but for the exploration of dust grain properties, multi-band measurements are necessary \citep{serkowski1973.polarization.spectrum.IAUS,whittet_1992.variations.wavelength.polarization,andersson_potter2010.maximum.alignment, cotton_2019.multi.wavelength.local.bubble}.

Variations in the degree of optical dust-induced polarization ($p_\lambda$) with wavelength (polarization spectrum) follow an empirical relation, which is usually referred to as the Serkowski relation \citep{serkowski1973.polarization.spectrum.IAUS, serkowski_1975}. The Serkowski relation provides important constraints on grain properties \citep{hensley_draine2021.dust.grains.constraints,hensley_draine_2023.pah.astrodust}. It reads as follows:
\begin{equation}
	\label{eq:serkowski_relation}
	p_{\lambda} = p_{max}~\exp \left[ - K ~ \ln^{2}{\left( \frac{\lambda_{max}}{\lambda} \right)} \right],
\end{equation} 
where \pmax\ is the maximum polarization fraction, usually observed at the V-band; \lmax\ is the wavelength where \pmax\ is observed, and is proportional to the average size of aligned dust grains \citep{mathis_1986.ISM.grain.alignment.superparamagnetic}; $K$ quantifies the spread of $p_\lambda$; large $K$ corresponds to narrow profiles, and vice versa. $K$ is considered a proxy for the dust grain size distribution \citep{serkowski_1975}, or for the optical property variations of silicate grains \citep{papoular_2018.interpretation.serkoski.curve}. 

Initially, it was found that $K = 1.15$ \citep{serkowski1973.polarization.spectrum.IAUS}, but more data suggested a linear relationship between $K$ and \lmax\ \citep{wilking_1980.linear.relation.K.lmax, whittet_1992.variations.wavelength.polarization, whitted_2022.ISM.dust.book},
\begin{equation}
	\label{eq:K_lambda_max_relation}
	K = (1.66 \pm 0.09) ~\lambda_{max} + (0.01 \pm 0.05).
\end{equation}
This is known as the Wilking relation. Eqs.~(\ref{eq:serkowski_relation}), and (\ref{eq:K_lambda_max_relation}) hold for $\lambda~\epsilon~[0.1, 1]$ \mum.   

The wavelength dependence of the degree of polarization can be understood as a consequence of the underlying (aligned) grain size distribution. Grain (symmetric and asymmetric) size distributions can be accurately approximated as power laws with negative slopes, indicating that small grains are more abundant than large grains \citep[MRN,][]{mathis_1977.MRN.distribution,weingartner_draine2001.grain.size.distribution}. Small grains (nanoparticles) are not easily aligned due to higher collision rates with gas particles \citep{hoang_laz_andersson_2015.rats.H2.formation}. Large grains are less susceptible to collisional disalignment but are less abundant. The two effects are balanced for intermediate-sized grains because they are neither too small nor too scarce. Consequently, the maximum degree of polarization is expected for intermediate size grains. In our Galaxy, this optimal size of aligned grains, which is proportional to \lmax, corresponds to \lmax~$\approx$ 0.55 \mum\ \citep{whitted_2022.ISM.dust.book}.
 
The empirical discovery of the polarization fraction's wavelength dependence by \cite{serkowski_1975} revolutionized multi-wavelength polarization studies; it provided a critical framework for understanding ISM dust and magnetic fields. Soon after this discovery, it became apparent that multi-band polarization is a powerful tool for studying ISM dust and magnetic fields. The impact of the 3D ISM structure on the Serkowski relation was predicted early on \citep{clarke_alroubaie_1984.multiple.clouds.affect.K.serkowski}, but observational evidence was limited due to the lack of data. In the last two decades, however, the influx of optical polarization data greatly increased, allowing for systematic explorations of the 3D effects on the $p_\lambda$ relation. This is evident by the recent agglomeration of starlight polarization data \citep{panopoulou_2025.polarization.catalogue}, which increased the number of polarization measurements by a factor of five from the previous agglomeration \cite{heiles_2000.polarization.agglomeration}.

\cite{mandarakas_2025} performed a multi-wavelength starlight polarization study to investigate the impact of the 3D ISM structure, and magnetic field morphology on the Serkowski fits. Through a combination of polarization data, Gaia distances, and 3D dust extinction maps \citep{edenhofer_2024.3d.extinction.map}, \cite{mandarakas_2025} showed that when multiple clouds along the line of sight (LOS) with different magnetic field geometries, or dust properties contribute to the observed polarization signal, $p_\lambda$ can either significantly deviate from or accurately follow the Serkowski relation. The latter case is particularly important: it highlights that the standard practice of only fitting the polarization measurements in the $p_\lambda$ - $\lambda$ space can be misleading due to the 3D structure of the ISM, even when the fits are accurate. In these cases, variations in the electric vector position angles (EVPA) should be taken into account through a joint fit in the Stokes $Q$ and $U$ space. This result has important consequences for the interpretations of multi-wavelength polarization data, and limits our capacity to probe the intrinsic properties of (aligned) dust. 

In light of this discovery, we assess how reliably Serkowski parameters can be inferred from starlight polarization observations and interpreted in terms of dust properties \citep{hensley_draine_2023.pah.astrodust} and grain alignment \citep{lazarian_hoang_2007.rat.theory,andersson_2015.review}. Interstellar dust studies based on optical polarimetry often impose the Wilking relation or a constant $K$ to compensate for limited wavelength coverage \citep[e.g.,][]{andersson_2007, panopoulou_2019_p_extinction,vaillancourt_2020.dust.grain.growth.using.multiband.polarization.data}. Instead, we adopt a probabilistic framework that accommodates both informed priors, which include the Wilking relation, and uninformed priors. We investigate how priors and LOS integration may affect the inferred parameters and their interpretation, focusing on the $K - \lambda_{max}$ relation and the polarization efficiency, $p_{max}/E(B-V)$.

The paper is organized as follows: In Sect.~\ref{sec:polarization_data} we present the employed dataset. In Sect.~\ref{sec:data_fitting}, we fit the Serkowski relation to the data using two methods. In Sect.~\ref{sec:wilking_relation}, we explore the correlations between $K$ and \lmax, and we present individual cases of measurements deviating from the Wilking relation. In Sect.~\ref{sec:polarization_efficiency}, we investigate the polarization efficiency properties, while in Sect.~\ref{sec:conclusions_future_prospects} we summarize the main outcomes of this work.

\begin{figure*}
	\centering
	\includegraphics[width=0.88\hsize]{Figures/serkowski_fits_characteristic_revised.pdf}
	\caption{Characteristic examples of fits with different EVPA variability metrics ($\Delta \theta$). Left: polarization degree as a function of wavelength. Right: EVPA as a function of wavelength. Blue curves show the Serkowski model evaluated at the marginal posterior medians obtained with uniform priors, while orange curves correspond to models obtained with informed priors. The legends give the obtained median parameters of the Serwkoski curve, including $\chi^2_p$. The orange and blue dotted vertical lines mark the median \lmax\ obtained with informed and uniform priors respectively. StarID, Rawstar, and refID correspond to the measurements' unique identifier and reference in the compiled catalogue.}
         \label{fig:serkowski_fits_examples}
\end{figure*}

\begin{figure}
	\centering
	\includegraphics[width=\hsize]{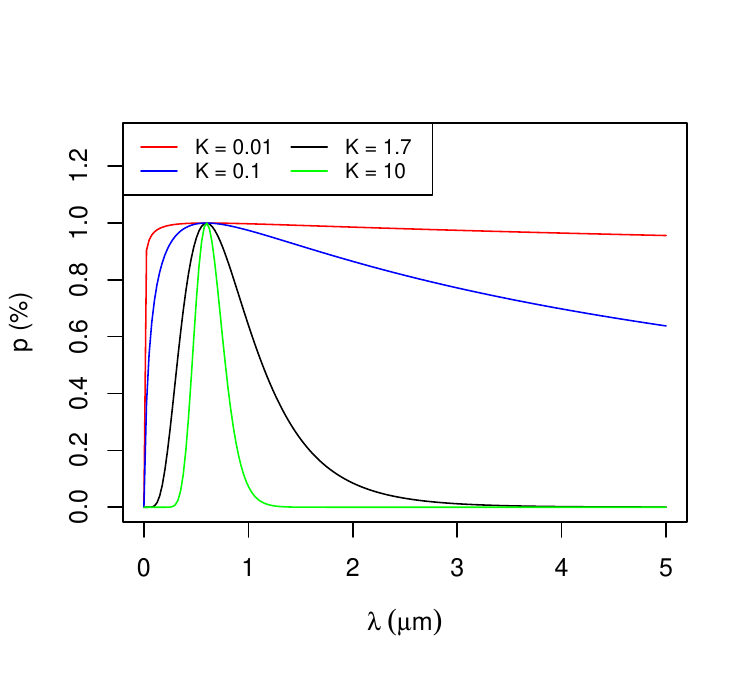}
	\caption{Analytical examples of the Serkowski relation (Eq.~\ref{eq:serkowski_relation}). The colored curves have different $K$ but fixed \pmax, and \lmax\ at $1\%$ and 0.6~\mum\ respectively. The black curve is typical of the ISM, while other curves, such as red, and green, are extreme cases that are unlikely to be observed in our Galaxy. Fitting models with only a few data around \lmax, leads to a spectrum of $K$ solutions, which can favor extreme values. For these cases, propagating prior knowledge was crucial for the fits.}
         \label{fig:serkowski_fits_example}
\end{figure}

\section{The multi-wavelength polarization sample}
\label{sec:polarization_data}

We extracted stars with polarization measurements in more than two bands from the compiled catalogue of \cite{panopoulou_2025.polarization.catalogue}. 
The polarization measurements used in our analysis were originally published in the following studies \citep{
whittet_2001.taurus.dust.polarization, 
oudmaijer_2001.export.stellar.polarimetry, 
fossati_2007.fors1.polarization.standards, 
weitenbeck_2008.ngc1502.polarization, 
weitenbeck_2008.hpol.measurements, 
medhi_2008.ngc654.polarization, 
medhi_2010.polarization.NGC6823,
andersson_potter2010.maximum.alignment, 
andersson_2013.ic63.grain.alignment,
eswaraiah_2011.ngc1893.polarimetry, 
eswaraiah_2012.polarimetry.young.open.cluster, 
eswaraiah_2013.ldn1570.polarimetry, 
santos_2014.sh2.29.polarization, 
santos_2012.rcw41.polarimetry, 
seronnavarrete_2016.rcw121.polarimetry, 
topasna_2018.ngc1502.extinction.polarization, 
topasna_2020.ngc6530.extinction, 
topasna_2021.ngc7160.extinction.polarization, 
topasna_2022,
neha_2018.mbm.clouds.polarization, vaillancourt_2020.dust.grain.growth.using.multiband.polarization.data, 
singh_2020.ngc1817.polarization, 
singh_2020.casado.alessi1.polarization, 
singh_2022.ngc2345.grain.alignment}.

Some stars contain measurements obtained in the same band at different epochs. For these cases, we calculated the average relative Stokes parameters ($\bar{q}$, $\bar{u}$) weighted by their observational uncertainties and then the weighted degree of polarization as $\bar{p} = \sqrt{\bar{q}^{2} + \bar{u}^{2}}$. The observational uncertainty of $\bar{p}$ is \citep{king_2014}
\begin{equation}
	\label{eq:sigma_p_weighted}
	\sigma_{p} = \sqrt{  \frac{\left (\bar{q}\sigma_{\bar{q}} \right )^2 +  \left( \bar{u}\sigma_{\bar{u}} \right)^2  }{\left (\bar{q}^2 + \bar{u}^2 \right)}  }
\end{equation}
where $\sigma_{\bar{q}}$, and $\sigma_{\bar{u}}$ correspond to the standard errors of $\bar{q}$ and $\bar{u}$ respectively. 

The reported observational uncertainties of some stars were very small, $\sigma_{p} \ll 0.1 (\%)$, but as we argue below, they are likely underestimated. The major challenge in achieving such low $\sigma_{p}$ is the instrumental calibration. Even if the instrumental polarization were zero, obtaining observational uncertainties smaller than 0.1$\%$ would be inevitable because of the lack of low-uncertainty polarimetric standard stars. 

Until recently, the availability of polarization standard stars was limited, while several stars were proven to be variable at the level of $0.1\%$ \citep[e.g.,][]{blinov_robopol_archive,blinov_2023.polarization.catalogue} or had wrong measurements \citep[e.g.,][]{skalidis_2018}. A robust polarimetric standard catalogue, which contains 65 stars mostly in the Northern hemisphere, was recently established \citep{blinov_2023.polarization.catalogue} \footnote{This four-years-long survey was performed with the RoboPol polarimeter \citep{robopol_paper_2019} mounted at the 1.3 m telescope at the Skinakas observatory in Crete, Greece.}. This is the most accurate catalogue of optopolarimetric standard stars to date with uncertainties $\sim 0.1 \%$, and sets the limit of instrumental calibration. For this reason, we uniformly set polarization uncertainties lower than 0.1 $\%$ to 0.1 $\%$. Be this as it may, our approach is conservative and leads to overestimating the uncertainties in the obtained fits.

\subsection{Variability in the position angles}
\label{sec:evpa_variability}

For several stars, EVPAs varied with the wavelength. Variations in EVPA are expected when: 1) different polarization mechanisms contribute to the observed signal \citep{lopez-rodriguez_2016.polarization.bayesian}; 2) the observed star has an intrinsic polarization mechanism, such as circumstellar disk, or it is photometrically variable; 3) the magnetic field and dust properties significantly vary along the LOS, hence the polarization signal traces the cumulative effect of the various LOS contributing structures \citep{patat_2010.vlt.polarimetry.NGC300,tassis_2015.frequency.decorrelation, guillet_2018.dust_models.Planck.compatibility,mandarakas_2025}.  

EVPA variability whether induced by the ISM or other factors, such as those described above, raises questions about the nature of the polarization signal, and how accurately it traces the dust grain properties. We identified measurements with variable EVPAs and treated them separately in our analysis. EVPA variability is quantified as:
\begin{equation}
	\Delta \theta \equiv \frac{\max(\theta_\lambda) - \min(\theta_\lambda)}{\sqrt{\sigma_{\max(\theta_\lambda) }^2 + \sigma_{\min(\theta_\lambda) }^2 } },
\end{equation}
where $\max(\theta_\lambda)$, and $\min(\theta_\lambda)$ correspond to the maximum and minimum EVPA across the various wavelength bands, while $\sigma$ denotes the corresponding observational uncertainties \footnote{Uncertainties in EVPAs depend on the S/N of the measurements and they are directly provided in the catalogue of \cite{panopoulou_2025.polarization.catalogue}. Averaging measurements obtained at different epochs, and increasing their observational uncertainties in $p$ (Sect.~\ref{sec:polarization_data}), decreases the S/N in $p$, hence increases the corresponding EVPA uncertainties. For these cases, we calculated the uncertainties using the EVPA intrinsic probability density function method \citep{Naghizadeh_1993.evpa.intrinsic.pdf}.}.

Following \cite{mandarakas_2025}, we distinguish measurements with constant and variable EVPAs. When \EVPAvariability~$\leq 1$, any maximum difference in the position angles is consistent within one sigma with the observational uncertainties, and in this case we can confidently consider that EVPAs remain statistically constant across wavelengths. On the other hand, when \EVPAvariability~$> 3$, variations in the position angles are statistically significant. We also considered measurements with intermediated variability, $1 < \Delta \theta \leq 3$. We considered a flag provided by \cite{panopoulou_2025.polarization.catalogue} to exclude intrinsically polarized sources. The full fitting sample contains 1,180 polarization spectra.\footnote{Companion catalogue containing the derived Serkowski parameters:
\url{https://doi.org/10.5281/zenodo.24376980}.}
 
\begin{figure*}
	\centering
	\includegraphics[width=\hsize]{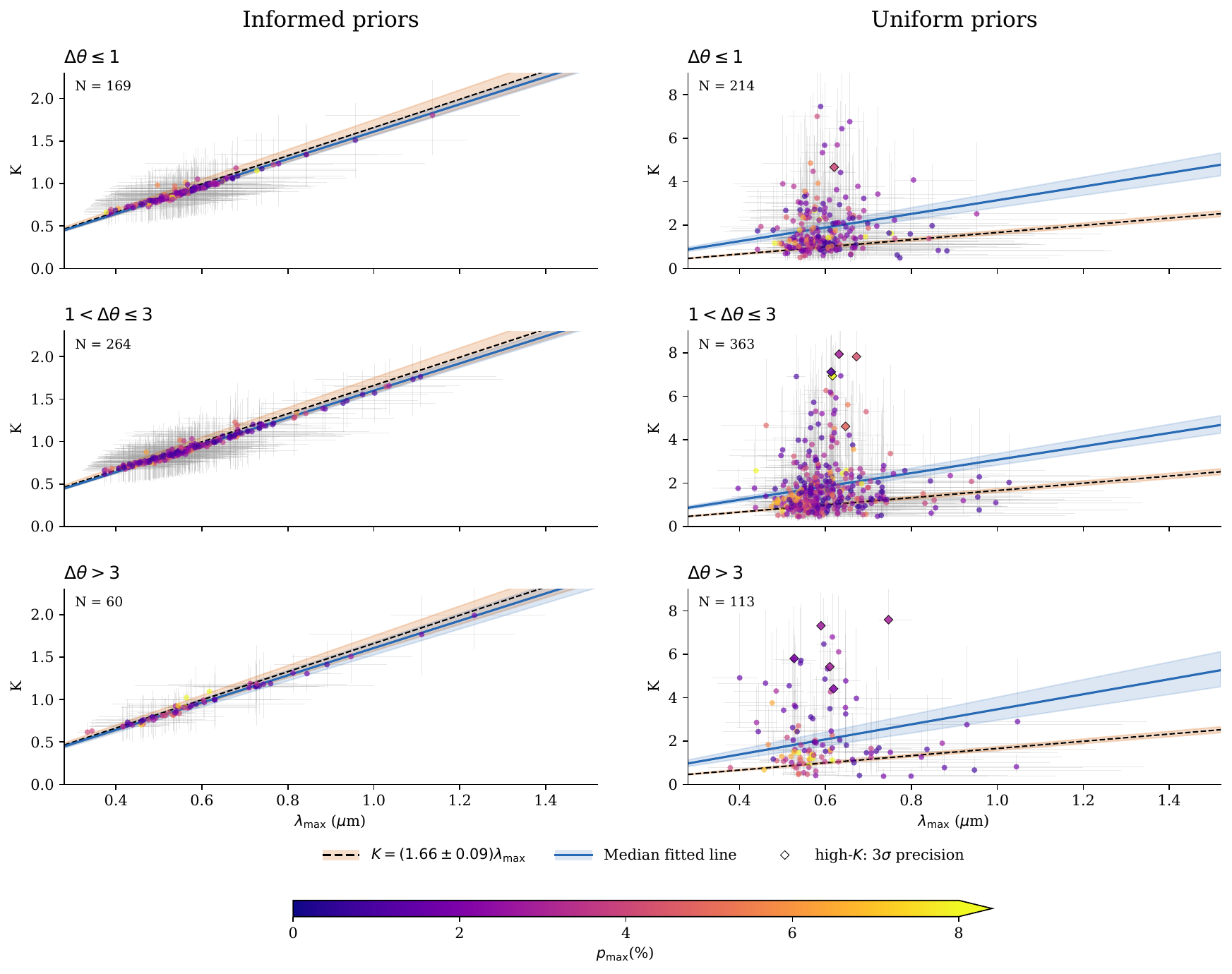}
	\caption{Comparison of $K$ and \lmax\ obtained with informed (left) and uniform (right) priors. From top to bottom, the rows correspond to $\Delta \theta \leq 1$, $1 < \Delta \theta \leq 3$, and $\Delta \theta > 3$. The selected number of measurements is shown in the legend of each panel. Points correspond to the posterior medians, while error bars span the 16th–84th percentile intervals of the posterior distributions. Colors represent the posterior median \pmax. The dashed line shows the reference relation (Eq.~\ref{eq:K_lambda_max_relation}). The blue line and shaded region show the median and central $68\%$ range of the fitted relations. Diamonds show stars with high $K$ values.
	}	
	\label{fig:wilking_relation_priors}
\end{figure*}
 
\section{Data fitting}
\label{sec:data_fitting}

We fit the multi-wavelength polarization data with two different ways: 1) in the $p - \lambda$, and 2) $q - u$ space. The latter was proposed by \cite{mandarakas_2025} to account for EVPA variability. 

\subsection{$p - \lambda$ fitting}
\label{sec:p_lambda_fitting}

The Serkowski relation (Eq.~\ref{eq:serkowski_relation}) is characterized by three free parameters. Fitting for these parameters requires measurements in at least three bands. Polarization studies, however, are usually limited to single-band observations because of observing time limitations. Even when multi-band measurements are obtained, the number of observed bands barely exceeds three and in addition, if these bands are obtained with the same instrument, their wavelength separation is not sufficient to accurately trace the $p_\lambda$ curves. This is a limitation of the existing dataset, which is the largest to date, that we cannot overcome. 

We employed a Bayesian framework to facilitate the fitting process from prior knowledge, and obtain the family of all solutions. This way we can also account for statistical biases that are inherent in low signal-to-noise (S/N) measurements, which have asymmetric confidence intervals in $p$ \citep{simmons_stewart_1985.polarization.debias,Plaszczynski2014.polarization.debiasing.method}. 

The likelihood, $\mathcal{L}(p_{obs} \mid \vec{\chi})$, is the probability density of obtaining the observed polarization measurements, $p_{obs}$, given the model parameters $\vec{\chi}$. The model polarization at each wavelength, $p_{\lambda}(\vec{\chi})$, is described by the Serkowski relation (Eq.~\ref{eq:serkowski_relation}), where $\vec{\chi}\equiv(p_{max}, K, \lambda_{max})$ is the vector of free parameters. Bayes' theorem gives the posterior distribution of these parameters as
\begin{equation}
    \label{eq:bayes_theorem}
    P(\vec{\chi}\mid\mathbf{p}_{\rm obs})
    \propto
    \mathcal{L}_p(\mathbf{p}_{\rm obs}\mid\vec{\chi})\,
    \pi(\vec{\chi}),
\end{equation}
where $\pi(\vec{\chi})$ is the prior distribution. We define a physically motivated likelihood and prior and use them to infer the posterior distribution of $\vec{\chi}$. For computational efficiency, we work with their logarithms:
\begin{equation}
    \label{eq:Bayes_theorem_log}
    \log P(\vec{\chi}\mid p_{obs})
    \propto
    \log\mathcal{L}_p (p_{obs}\mid\vec{\chi})
    + \log\pi(\vec{\chi}) .
\end{equation}

\begin{figure}
	\centering
	\includegraphics[width=\hsize]{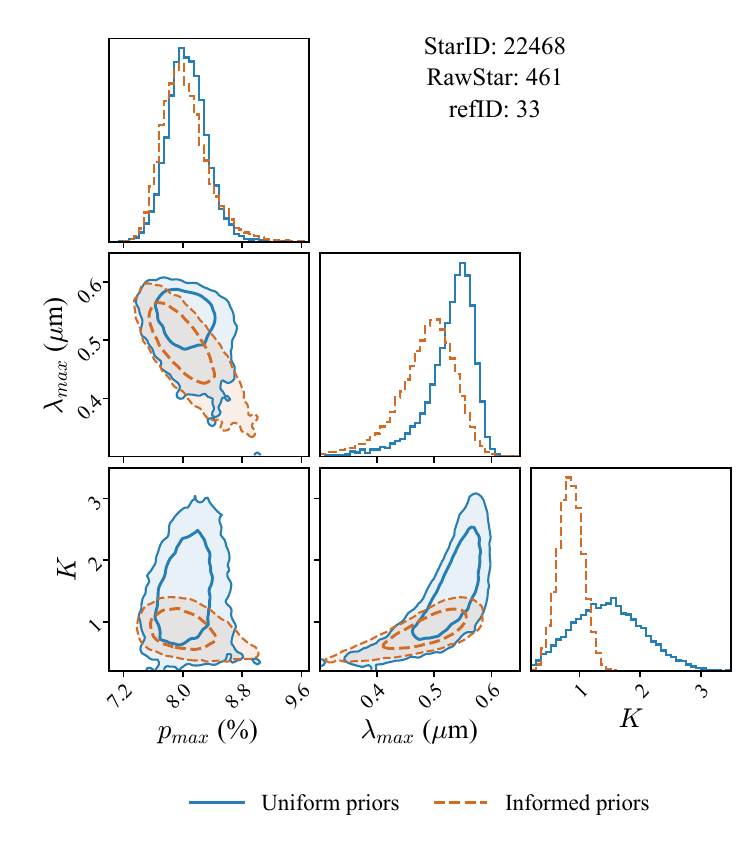}
	\caption{Posterior distributions of the Serkowski parameters \pmax, $K$, and \lmax\ for the star with Rawstar 461, whose polarization fits are shown in Fig.~\ref{fig:serkowski_fits_examples}. Blue solid and orange dashed contours correspond to the uniform- and informed-prior fits respectively. Contours enclose $68\%$ and $95\%$ of the posterior probability.}
        \label{fig:posteriors_correlation}
\end{figure}

\subsubsection{Definition of likelihood}
\label{sec:likelihood_definition}

Gaussian likelihoods are usually employed in Bayesian statistics; this implies symmetric confidence intervals. Polarization fraction, however, is a positive definite quantity, and follows the Rice distribution, which for low S/N is asymmetric, bounded at zero S/N, and has long tails to large values \citep{rice_1945.rice.distribution}. The Rice distribution approximates normal for S/N $\gtrsim 5$ (Appendix~\ref{appendix:likelihoods}). A large fraction of the employed data has a low S/N ratio; for these cases, we need to consider non-Gaussianities   \citep{simmons_stewart_1985.polarization.debias,vaillancourt_2006,Plaszczynski2014.polarization.debiasing.method}.

The probability, $\mathcal{L}_p (p_{obs} | p_{\lambda}(\vec{\chi}))$, of observing $p_\lambda$ is \citep{vaillancourt_2006}:
\begin{equation}
	\label{eq:likelihood}
	\mathcal{L}_p  = \Pi_{j=1}^{N} \frac{1}{N_{L, j}}\frac{p_{obs, j}}{\sigma_j^2}\exp{\left( \frac{- p_{\lambda}^2 - (p_{obs, j})^2}{2\sigma_j^2} \right)}~ \\
	I_0 \left( \frac{p_{obs, j} p_{\lambda}}{\sigma_j^2} \right). 
\end{equation} 
$I_{0}$ is the zeroth-order modified Bessel, and $j$ is the index that runs across the various wavelength bands. If, for example, we have $N=3$ polarization measurements (at different bands) then $j = \{1, 2, 3\}$. We retain the normalization factor $N_{L, j}$ following \cite{vaillancourt_2006},
\begin{equation}
	\label{eq:normalization_factor}
	N_{L, j} = p_{obs, j}\sqrt{\frac{\pi}{2}} \exp{\left( -\frac{p_{obs, j}^2}{4\sigma_j^2} \right)}~I_0\left(\frac{p_{obs, j}^2}{4\sigma_j^2} \right).
\end{equation}
Since $N_{L, j}$ depends only on observed quantities, it is independent of the fitted Serkowski parameters and cancels upon normalization of the posterior, and thus does not affect the inferred parameters.

Modified Bessel functions of the first kind, $I_n (x)$, grow exponentially, and as a result lead to numerical infinites for large $x$. This can be a major problem in numerical calculations, which become unstable for large $x$. More specifically, in our software -- we used the built-in functions of the R programming language to calculate $I_0$ -- we obtained infinities for $x >700$. This problem was more prominent for the Bessel function shown in Eq.~(\ref{eq:likelihood}) than in $N_{L, j}$ (Eq.~\ref{eq:normalization_factor}). To treat these infinities, we used  the asymptotic form of $I_0$, which is $I_{0}(x) \sim e^{x}/\sqrt{2\pi x}$ for $x \gg 1 $. In this case, the likelihood function becomes asymptotic to 
\begin{equation}
	\label{eq:likelihood_asymptotic}
	\mathcal{L}_p \sim \Pi_{j=1}^{N} \frac{1}{N_L}\sqrt{\frac{p_{obs, j}}{p_{\lambda}}}\frac{1}{\sigma_j \sqrt{2\pi}} \exp{\left( - \frac{ (p_{obs, j} - p_\lambda)^2}{2\sigma_j^2} \right)},
\end{equation} 
which is a Gaussian, with a modified amplitude, and has symmetric confidence intervals \citep{vaillancourt_2006}. The asymptotic form of the normalization factor is $N_{L,j} \sim \sigma_j$. The above expression is nearly identical to the Rice distribution for S/N $\geq 5$, and could be used for S/N$\geq 4$. It is more accurate than a normalized Gaussian (Appendix~\ref{appendix:likelihoods}), and its simplicity makes it useful for future studies.

\subsubsection{Distribution of priors}
\label{sec:prior_distributions}

The maximum $p$ value of our extracted sample is $33\%$. For this reason, we assumed a uniform prior distribution for \pmax~($\%$) in the range [0, 33.0]; \lmax\ ranges from 0.3 to 1.0 \mum\ in our Galaxy \citep{wilking_1980.linear.relation.K.lmax, whitted_2022.ISM.dust.book}. To be conservative, we considered a uniform distribution for \lmax\ with slightly wider limits: $\lambda~\epsilon~[0.2, 1.5]$ \mum.

The employed data usually include a few measurements around the peak of the polarization curve, $\lambda \approx$ \lmax. This is not a problem for constraining \pmax, and \lmax, but $K$ becomes challenging. Without measurements at the near- ultraviolet (UV), and infrared (IR) wavelengths the shape of the polarization spectrum relation can be weakly constrained. Thus, S/N ratios in the obtained $K$ solutions are in general lower than the other two parameters. This problem is better understood by examining some extreme cases.

In the $K \rightarrow 0$ limit, $p \approx p_{max}$ for any $\lambda$, meaning that $p$ is independent of $\lambda$  (red curve in Fig.~\ref{fig:serkowski_fits_example}). If only a few measurements were considered around \lmax, then $K \approx 0$ solutions would be favored. The constancy of $p_\lambda$ would imply that grain alignment efficiency is independent of the grain size, which is unphysical. In the extreme case where $K \gg 1$, $p_{\lambda}$ approximates a delta function because it rapidly goes to zero everywhere but \lmax, where $p_{\lambda_{max}} \approx p_{max}$ (green curve in Fig.~\ref{fig:serkowski_fits_example}); this implies that there is a very narrow range of sizes of dust grains contributing to dust polarization. Neither of these two extreme cases is expected in nature \citep{whitted_2022.ISM.dust.book}. The lack of near-UV, and near-IR measurements in the fitting process favors the aforementioned unphysical cases. For this reason, considering a more stringent prior distribution for $K$ is necessary for our analysis.  

A linear correlation applies between $K$ and \lmax\ with a slope close to 1.66 \citep{wilking_1980.linear.relation.K.lmax}, when optical polarization data are used to fit the Serkowski relation. When UV polarimetric data are included in the fits, the slope between the two quantities increases: $K = (2.56 \pm 0.38)~\lambda_{max} + (-0.59 \pm 0.21)$ \citep{marting_1999.uv.polarization.spectrum}. In both cases the intercept is consistent with zero, but the difference in the slopes is statistically significant; likely induced by the peculiar behavior of dust polarization at UV wavelengths. Differences in the slopes obtained from fits with or without the UV data, become prominent only for \lmax\ < 0.5 \mum\ \citep{whitted_2022.ISM.dust.book}. This highlights the importance of obtaining UV polarization measurements \citep{anderssonBG_2022.uv.polarization.mission}, but observational constraints are limited there. For this reason, we shall restrict ourselves to constraints obtained from optical data (Eq.~\ref{eq:K_lambda_max_relation}).

We assumed that $K$ linearly correlates with \lmax\ as: $K = \alpha~\lambda_{max}$. We assumed that $\alpha$ follows a Gaussian distribution with an average value equal to 1.66, as the optical data suggests \citep{wilking_1980.linear.relation.K.lmax,whittet_1992.variations.wavelength.polarization,whitted_2022.ISM.dust.book}. Intrinsic variations about this slope were determined to be $5\%$, but, to be conservative, we increased the variance to $20 \%$. The slope between $K$ and \lmax\ is positive definite, $\alpha \geq 0$. Thus, we truncated the Gaussian distribution of $\alpha$ for $\alpha < 0$. We only considered measurements with S/N ratio in $p$ equal or larger than three. We also fit the spectra using independent uniform priors on all three parameters.

\subsection{$q - u$ fitting}
\label{sec:stokes_qu_fitting}

Constraining the Serkowski parameters through the $q - u$ space is an elaborate way to account for both $p_\lambda$, and $\theta_\lambda$ variations. Contrary to the degree of polarization, Stokes $q$, and $u$ are characterized by Gaussian likelihoods, which significantly simplifies the calculations. However, only recently was this method applied to multi-wavelength starlight polarization data \citep{mandarakas_2025}. Henceforth, we follow this methodology.

Stokes $q$, and $u$ at a given wavelength are:
\begin{align}
    	q_\lambda = p_\lambda~\cos(2\theta_\lambda), \\
    	u_\lambda = p_\lambda~\sin(2\theta_\lambda).
\end{align}
Substituting the Serkowski formula (Eq.~\ref{eq:serkowski_relation}) to the above equations yields
\begin{align}
	q_\lambda = p_{max} ~ \exp \left[-K ~ \ln^2\left({\frac{\lambda_{max}}{\lambda}} \right) \right] ~ \cos(2\theta), \\
	u_\lambda = p_{max} ~ \exp \left[-K ~ \ln^2\left({\frac{\lambda_{max}}{\lambda}} \right)  \right] ~ \sin(2\theta).
\end{align}

The product of the likelihoods of Stokes $q$, and $u$ is a 2D Gaussian that represents the probability of observing a measurement in the $q - u$ space. We assumed that Stokes $q$, and $u$ measurements are uncorrelated, hence no cross-talk between the two terms exists in their covariance matrix. The joint likelihood, $\mathcal{L}_{q, u} \equiv \mathcal{L} \left (q_{obs}, u_{obs}~|~q_{\lambda}(\vec{\chi}), u_{\lambda}(\vec{\chi}) \right )$, becomes:
\begin{align}
	\mathcal{L}_{q, u} = & \Pi_{j=1}^{N} \frac{1}{2\pi\sigma_{q, j}~\sigma_{u, j}} \exp{\left( -\frac{\left( q_{\lambda} - (q_{obs, j})\right)^2}{2\sigma_{q, j}^2} -\frac{\left(u_{\lambda} - (u_{obs, j})\right)^2}{2\sigma_{u, j}^2}\right)} 
\end{align}
where symbols have the same meaning as in Eq.~(\ref{eq:likelihood}). We maximized the logarithms of the above likelihood (Eq.~\ref{eq:Bayes_theorem_log}) to fit the data. We used the same prior distributions as in the $p - \lambda$ fits (Sect.~\ref{sec:prior_distributions}). The two fitting methods (Sects.~\ref{sec:p_lambda_fitting} and \ref{sec:stokes_qu_fitting}) yielded consistent results.

\subsection{Likelihood maximization and obtained fits}
\label{sec:likelihood_marginalization}

We maximized the log-likelihood (Eq.~\ref{eq:Bayes_theorem_log}) using the "BayesianTools" library implemented in R \citep{BayesianTools}. We used the default sampler, {\it DEzs}, and initiated 100 walkers, each running for 1000 steps. To ensure the convergence of the walkers, we removed from the posteriors the first 400 steps of the walkers. 

We used the chi-square to assess the quality of the fits:
\begin{equation}
\chi^2_p = \Sigma_j \frac{(p_{\lambda} (\vec{\chi}) - p_{obs, j} )^2}{\sigma_{j}^{2}}.
\end{equation}
We flag the fits according to their $\chi^2_p$ values, using $\chi^2_p \leq 1$ as our selection threshold. Informed-prior fitting yields 493 objects with $\chi^2_p \leq 1$, whereas the uniform-prior fitting 690 measurements. Examples of fits satisfying this criterion are shown in Fig.~\ref{fig:serkowski_fits_examples}.

\begin{figure}
	\centering
	\includegraphics[width=0.7\hsize]{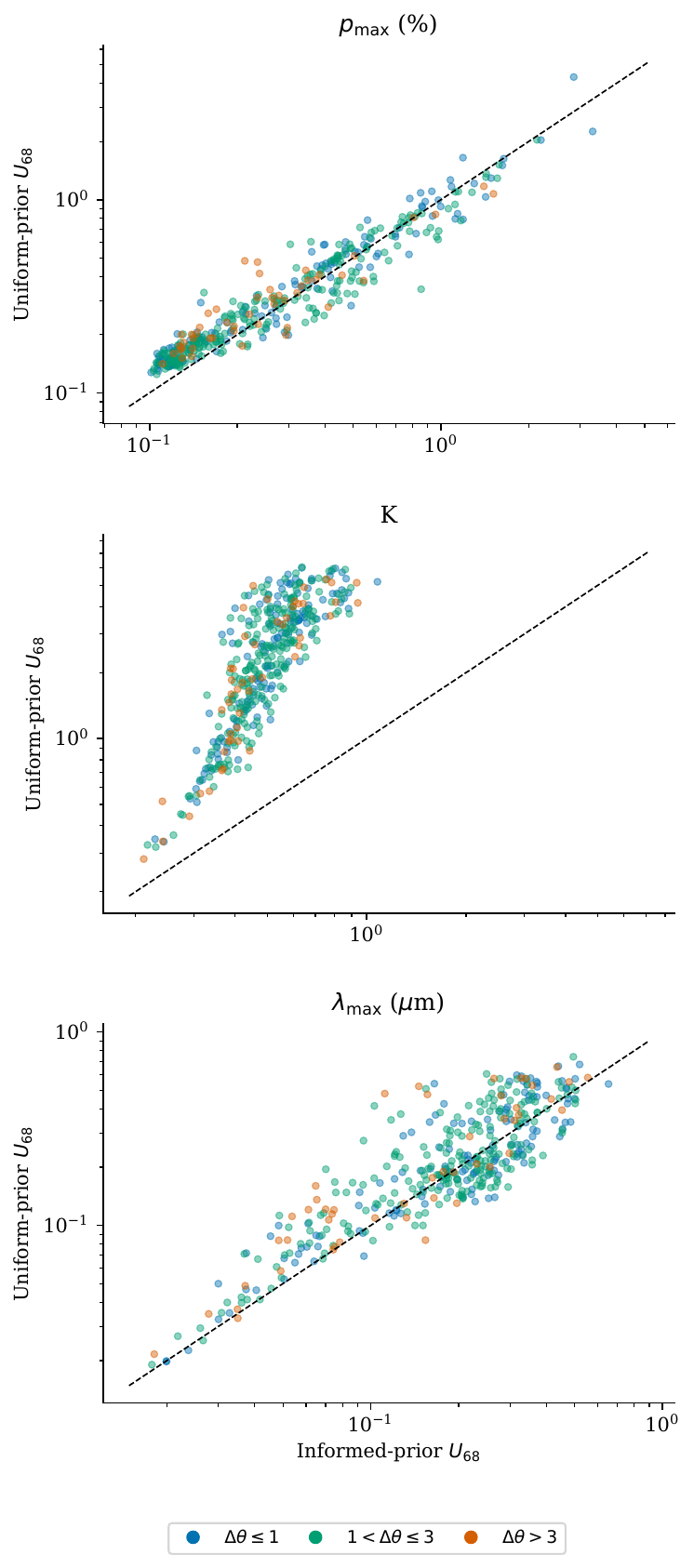}
	\caption{	Comparison of the posterior uncertainties in $p_{max}$, $K$, and \lmax\ obtained using informed and uniform priors. In total, 416 measurements have been selected with $\chi^2 \leq 1$ under both priors. Colors correspond to different $\Delta \theta$ groups as shown by the legend. Dashed lines indicate the one-to-one relation.
	}
	\label{fig:uncertainties_prior_dependence}
\end{figure}

\begin{figure*}
	\centering
	\includegraphics[width=0.7\hsize]{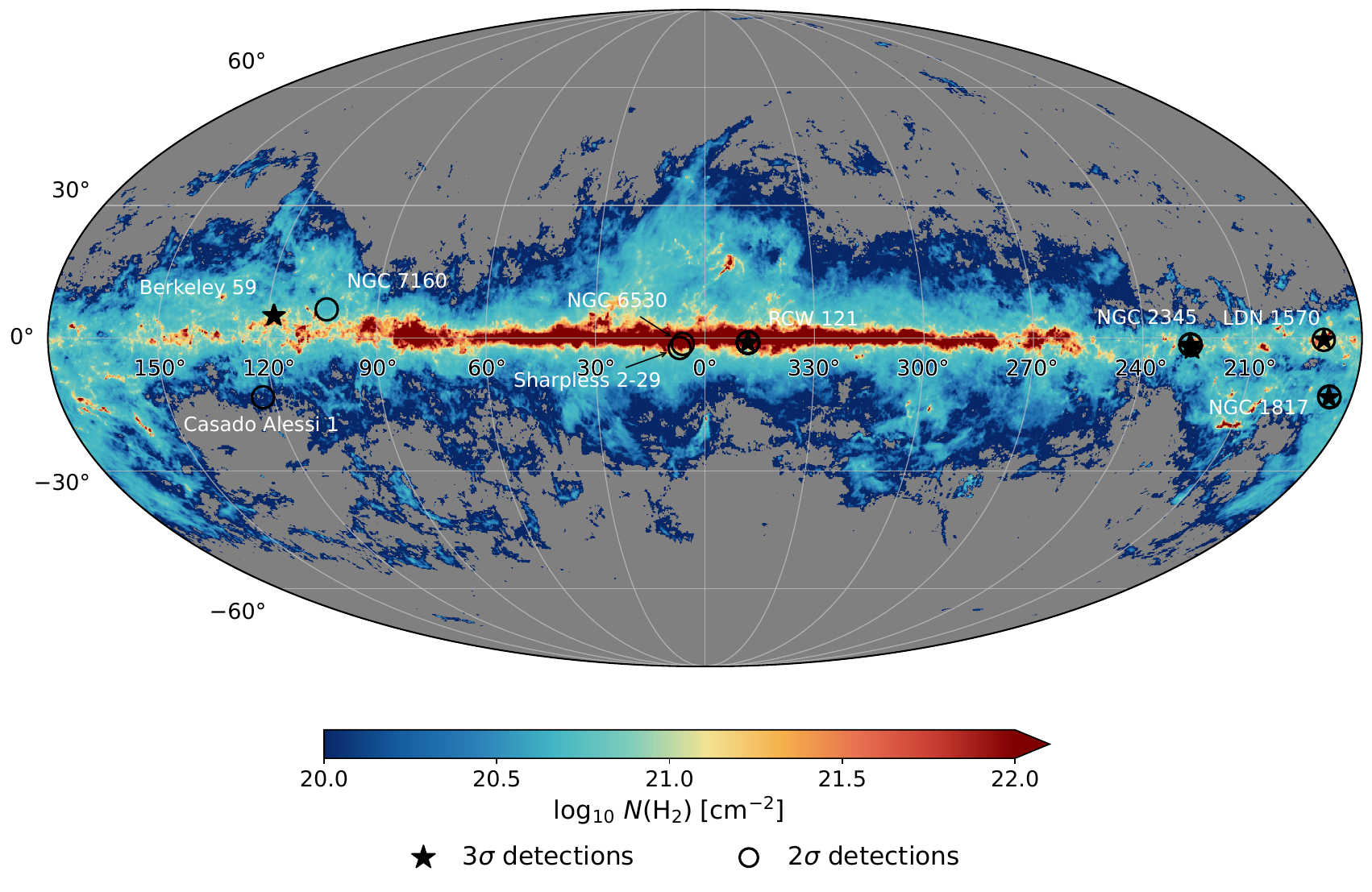}
	\caption{Sightlines of the high-$K$ measurements over plotted on an all-sky map of the molecular hydrogen column density in Galactic coordinates. Black stars indicate the 11 high-K measurements that satisfy the $3 \sigma$ precision criterion, while open circles show the 19 additional candidates that satisfy the $2 \sigma$ criterion.
	}
	\label{fig:NH2_map_highK}
\end{figure*}

\begin{figure*}
	\centering
	\includegraphics[width=0.9\hsize]{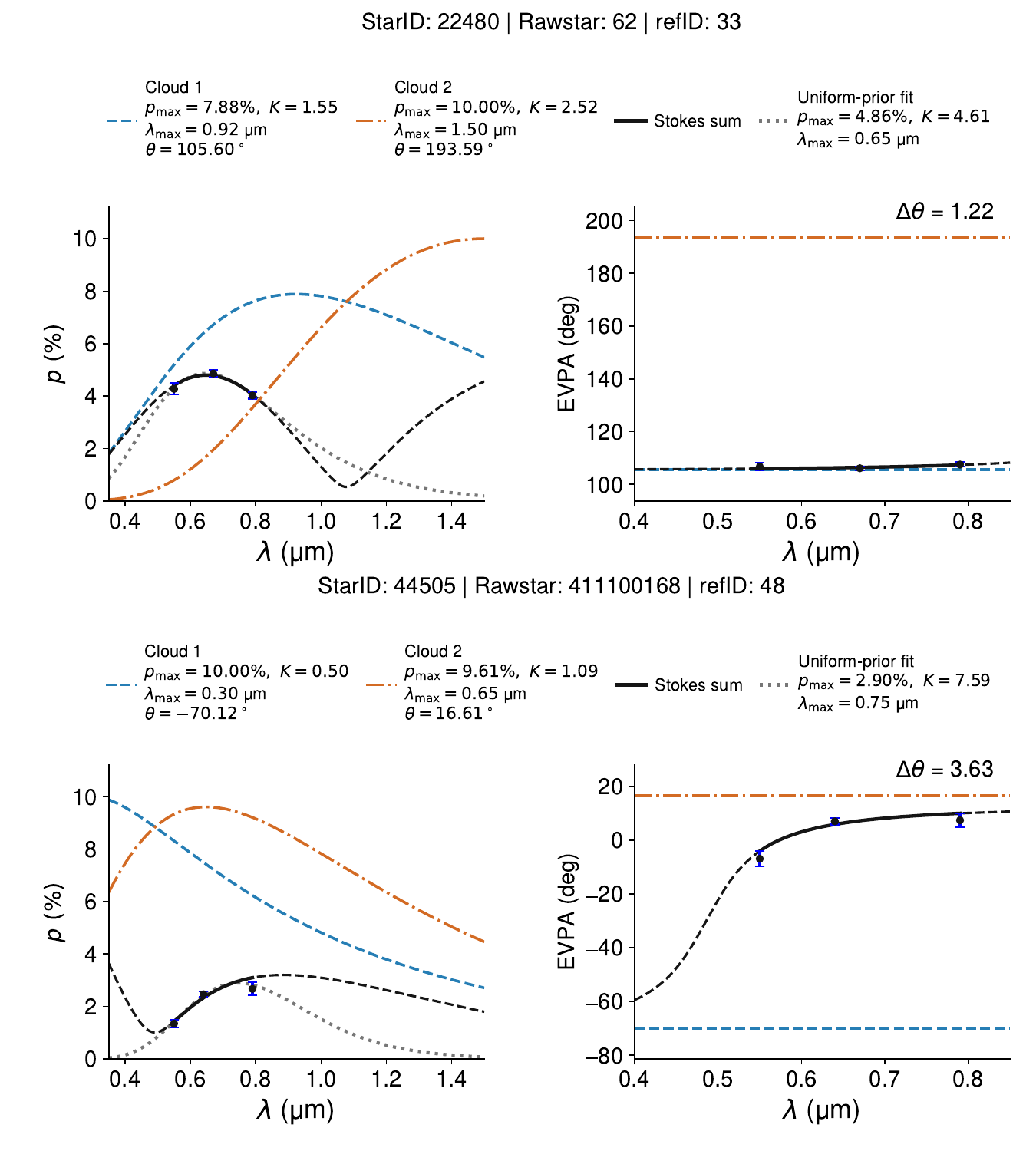}
	\caption{Two-cloud polarization models for StarID 22480 (top) and 44505 (bottom). Left: polarization degree as a function of wavelength. Right: EVPA as a function of wavelength. Black points with blue error bars show the observations, while the colored curves show the contributions of the individual clouds. Each cloud follows the Wilking relation, and its Serkowski parameters and EVPA are given in the legends. The solid black curves show the combined Stokes polarization, and the dotted curves show results from the uniform-prior fits. The parameters of these combined fits are given in the legends.}
	\label{fig:two_cloud_model}
\end{figure*}

\section{The Wilking relation}
\label{sec:wilking_relation}

The Wilking relation is commonly imposed when fitting polarization measurements with limited wavelength coverage to reduce the number of free parameters \citep[e.g.,][]{vergne_2018.multiband.pol.centaurus,singh_2022.ngc2345.grain.alignment}. However, imposing this relation also restricts the range of models allowed by the fit. We assessed how priors influence the inferred Serkowski parameters and whether the reference model (Eq.~\ref{eq:K_lambda_max_relation})  is supported by the existing data.

\subsection{Sensitivity of the Wilking relation to the prior distribution}

Fig.~\ref{fig:wilking_relation_priors} shows the posterior medians of $K$ and $\lambda_{\max}$ for measurements selected to have $\chi_p^2 \leq 1$. We apply the $\chi_p^2 \leq 1$ criterion separately to the informed- and uniform-prior fits in the left and right columns, respectively. We considered three groups with distinct EVPA variability metrics to evaluate the sensitivity of the results to LOS effects according to \cite{mandarakas_2025}. The number of measurements in each $\Delta \theta$ group is indicated in the legend of each panel. Error bars span the marginal 16th--84th percentile intervals, and colors show $p_{max}$ from the corresponding fit.

To determine if the data favor the Wilking relation, we fit a linear model to the data with zero intercept. We generate 2,000 realizations by drawing jointly from the $K$ and $\lambda_{\max}$ posterior distributions. The blue lines and shaded region show the median and central $68\%$ range of the resulting fitted relations.

The informed-prior results are consistent with the reference relation (black dashed line), as expected because the informed prior favors this coupling. Thus, this agreement does not by itself provide independent confirmation of the Wilking relation.

In contrast, the uniform-prior fits yield a broader distribution and steeper fitted relations in all three $\Delta \theta$ groups. Many measurements follow the Wilking relation with \lmax\ ranging from 0.4 to 1.0 \mum, while others cluster around \lmax~$\approx 0.6$ \mum\ but span a broad range of $K$. The higher-$K$ values in this distribution contribute to the steeper fitted slopes.

Some uniform-prior fits have \lmax~$> 0.8$ \mum\ values and are consistent with the reference relation. For some of these measurements, \lmax\ lies beyond the observed wavelength range and their posterior medians $K$ and \lmax\ are correlated. The middle row of Fig.~\ref{fig:serkowski_fits_examples} illustrates an example, where the observed polarization rises across the sampled bands, but the turnover is not constrained.

Limited wavelength coverage makes $K$ particularly difficult to constrain. Fig.~\ref{fig:posteriors_correlation} illustrates this sensitivity by comparing the joint posterior distributions of the Serkowski parameters for one measurement under uniform and informed priors. The marginal posterior for $K$ is broader under the uniform prior, and the posterior medians of both $K$ and \lmax\ shift when the informed prior is adopted.

To investigate the influence of priors on the obtained parameters across all measurements, we calculated the uncertainty for each obtained parameter as $U_{68}=x_{84}-x_{16}$, where $x_{16}$ and $x_{84}$ are the 16th and 84th percentiles of the marginal posterior distribution. Fig.~\ref{fig:uncertainties_prior_dependence} compares the posterior uncertainties obtained with the two prior distributions for overlapping measurements with $\chi^2_p \leq 1$. The dashed lines indicate the one-to-one relation. For \pmax\ and \lmax, most measurements lie close to this line, indicating comparable uncertainties for the two priors. In contrast, the uncertainties in $K$ are generally larger with uniform priors, indicating that the informed prior provides a stronger constraint on the spectral width. The sensitivity of $K$ to the adopted prior is present across all three $\Delta\theta$ groups. 

However, we also note informed priors can sometimes worsen the fits. For the high-$K$ measurements shown as diamonds in Fig.~\ref{fig:wilking_relation_priors}, the informed-prior fits yield $\chi^2_p>1$, whereas the corresponding uniform-prior fits yield $\chi^2_p\leq1$ (Table~\ref{tab:highK_measurements}). Their corresponding polarization spectra are shown in Fig.~\ref{fig:extreme_k_spectra}. These examples indicate that the Wilking relation is inconsistent with these measurements. We discuss them further in the following section.

\subsection{Deviations from the Wilking relation: high-$K$ measurements}
\label{sec:wilking_relation_deviations}

To identify high-$K$ candidates, we require $\chi^2_p \leq 1$, a posterior median $\operatorname{med}(K) \geq 3$, and $\operatorname{med}(K)$ to lie above the reference relation by more than five times the uncertainty in the predicted value arising from the quoted uncertainty in its slope. We also require $\operatorname{med}(x) \geq 3U_{68}(x)/2$ ($3\sigma$ detection) for both $x = K$ and $x=$ \lmax. We identified 11 measurements that satisfy the above criteria and are shown as diamonds in Fig.~\ref{fig:wilking_relation_priors}. The properties of these measurements are summarized in Appendix~\ref{appendix:high_k_stars}.

We overplot the sightlines toward the high-$K$ candidates on the molecular hydrogen column-density ($N_{\mathrm{H}_2}$) map of \citet{skalidis_2024.H2.column.densities} to show their projected locations relative to known molecular structures. The map also includes 19 additional candidates (shown as black circles) selected by relaxing the relative-precision requirements to $\operatorname{med}(x) \geq 2 U_{68}(x)/2$ ($2\sigma$ detection), while retaining the rest of the cuts. 
 
The high-$K$ candidates span nine observed fields: 1) six open clusters, including Berkeley 59, NGC 6530, NGC 1817, NGC 7160, NGC 2345, and Casado Alessi 1, 2) a dark cloud, LDN 1570, and 3) two H\,\textsc{ii} regions, including Sh 2-29 and RCW 121 (Table~\ref{tab:highK_measurements}). The Serkowski profiles of these measurements appear narrower than expected from the Wilking relation. Similar departures have also been reported along extragalactic sightlines toward Type Ia supernovae \citep{patat_2015.extragalactic.SN, cicota_2018.SN_polarization}.

The radiative torque disruption theory \citep{hoang_2019.RATD.theory} can produce elevated $K$ values by modifying the aligned grain-size distribution. However, this theory predicts blue-shifted wavelength peaks (\lmax~$< 0.5$ \mum) \citep{giang_2020.extinction.polarization.RATD}, while our measurements indicate red-shifted profiles (\lmax~$> 0.6$ \mum). The evolution dust models of \cite{voshchinnikov_2014.grain.growth.evolution.serkowski.curves} provide an alternative explanation. These authors find that more elongated oblate grains produce narrower polarization spectra in their models, although none of their presented modeled curves seem to reach values as high as $K = 8$. According to the dust modeling of \cite{cicota_2018.SN_polarization}, elevated $K$ and redshifted \lmax\ can occur when the population of large silicate grains is suppressed. This is consistent with our observed trends but whether this mechanism can account for the high-$K$ measurements found here remains to be investigated. 

LOS superposition offers an additional explanation for the large values of $K$. Fig.~\ref{fig:two_cloud_model} illustrates the polarization profiles of a two-cloud model constructed in the $q - u$ space following the formalism of \cite{mandarakas_2025}\footnote{\cite{martin_1974.polarized.transfer} provides a general treatment of polarized transfer through a medium in which the grain alignment direction varies along the LOS.}. We assumed that each cloud follows a Serkowski law satisfying the Wilking relation and has a wavelength-independent polarization angle $\theta_i$.  When the clouds have different $\lambda_{\max,i}$ and polarization angles, their relative contributions change with wavelength. This can modify the apparent width of the polarization profile, even though neither individual cloud departs from the reference $K$--$\lambda_{\max}$ relation.

For the stars shown in Fig.~\ref{fig:two_cloud_model}, the uniform-prior fit yield $(\lambda_{\max}, K)=(0.65\,\mu{\rm m}, 4.61)$ and $(0.75\,\mu{\rm m}, 7.59)$, respectively. Both profiles are substantially narrower than predicted by the reference relation at their fitted \lmax\ values. These examples illustrate that the superposition of cloud components that individually follow the Wilking relation can produce an apparently elevated $K$. This, however, does not exclude a contribution from variations in grain properties and alignment efficiency as discussed by \cite{voshchinnikov_2014.grain.growth.evolution.serkowski.curves}, \cite{giang_2020.extinction.polarization.RATD}, and \cite{cicota_2018.SN_polarization}. Additional near-infrared polarization measurements around $1\,\mu{\rm m}$ would help test the models, particularly where their predicted spectra diverge. 

\begin{figure*}
	\centering
	\includegraphics[width=0.9\hsize]{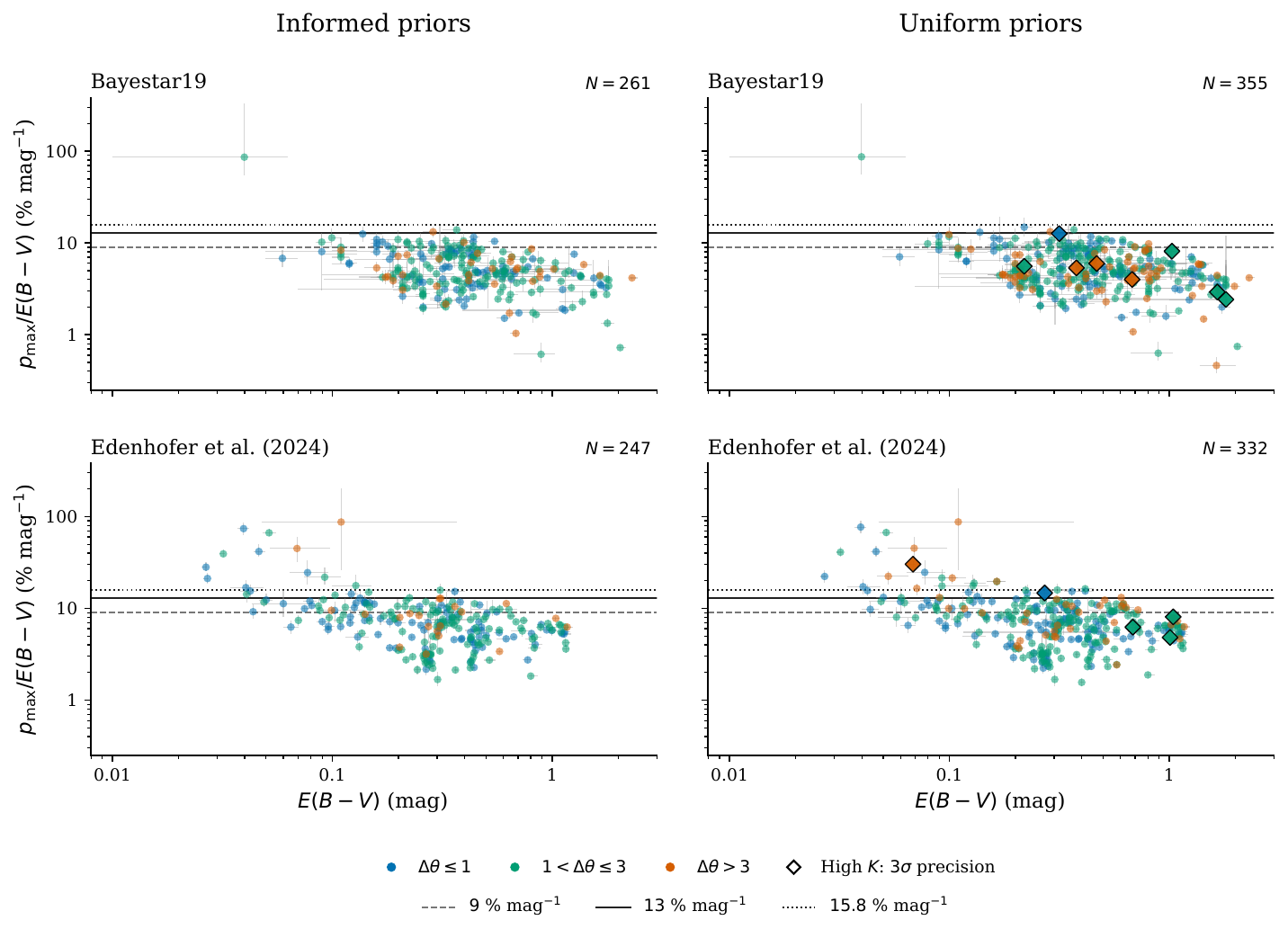}
	\caption{Polarization efficiency, $p_{max} / E(B-V)$, as a function of reddening for the informed-prior (left) and uniform-prior (right) fits. Reddening estimates are obtained from Bayestar19 (top) and \cite{edenhofer_2024.3d.extinction.map} (bottom). Blue, green, and orange circles represent measurements with $\Delta \theta \leq 1$, $1 < \Delta \theta \leq 3$, and $\Delta \theta > 3$ respectively. Diamonds identify the high-$K$ measurements. Horizontal black lines indicate the various polarization-efficiency limits as indicated in the legends.}
	\label{fig:pol_efficiency}
\end{figure*}

\section{Polarization efficiency}
\label{sec:polarization_efficiency}

Polarization efficiency, represented as $p_{max}$ / \EBV, quantifies the ability of aligned dust grains to polarize starlight and provides a crucial constraint for dust grain models \citep{hensley_draine2021.dust.grains.constraints}. Complex magnetic field topologies, especially when the field has a substantial component along the LOS, can reduce the observed \pmax\ values \citep[e.g.,][]{angarita_2023.polarization.efficiency}. Thus, this observable depends on both grain properties and magnetic field geometry.

The upper limit in $p$~/~\EBV\ was initially estimated around $9\,\%\,{\rm mag}^{-1}$ \citep{serkowski_1975}. However, in several diffuse ISM regions with \EBV~$\sim 0.01$ mag, this limit was systematically exceeded \citep{skalidis_2018,panopoulou_2019_p_extinction,angarita_2023.polarization.efficiency}, suggesting that the empirical upper envelope is higher than previously estimated. By combining polarized dust emission and optical polarization data, \cite{planck_2020.ISM.polarization.dust.physics} obtained an upper-envelope estimate of $p_V/E(B-V)\simeq13\,\%\,{\rm mag}^{-1}$. We note, however, that polarization efficiencies as high as $15.8\,\%\,{\rm mag}^{-1}$ have been reported in the diffuse ISM \citep{angarita_2023.polarization.efficiency}. \cite{panopoulou_2019_p_extinction} pointed to uncertainties associated with dust extinction maps that can significantly influence the obtained constraints.

We examined the polarization efficiencies in our sample. Fig.~\ref{fig:pol_efficiency} shows \pmax~/~\EBV\ versus \EBV\ for the informed-prior and uniform-prior fits in the left and right columns, respectively. The top row uses Bayestar19 \citep{green_2018, green_2019.bayestar}, while the bottom row uses the map of \cite{edenhofer_2024.3d.extinction.map}. \EBV\ is computed up to the distance of the star and stellar distances from Gaia \citep{gaia_dr3} using the geometric distances from \cite{bailer_jones_2021}. The colors identify the three $\Delta\theta$ groups, and diamonds mark the high-$K$ measurements. We display measurements with valid-distance availability, map coverage from each map, and \EBV~$\geq0.02$ mag as an approximate statistical threshold \citep[e.g.][]{panopoulou_2019_p_extinction}.

For Bayestar19 we use the median reddening, with $E(B-V)=0.996E_{\rm Bayestar}$. For the \citet{edenhofer_2024.3d.extinction.map} map, we integrate the posterior-mean differential extinction along each sightline to the stellar distance to obtain the cumulative extinction $E_{\rm ZGR}$ in the native map units. Following their conversion, we approximate the $V$-band extinction by $A_V \approx A(540\,{\rm nm}) = 2.8\,E_{\rm ZGR}$, with $A_V$ in magnitudes. Adopting $R_V \equiv A_V/E(B-V)=3.1$, we obtain $E(B-V) \approx (2.8/3.1)\,E_{\rm ZGR}$. In contrast to $K$ and $\lambda_{\max}$, we do not find significant differences between the resulting $p_{\max}/E(B-V)$ estimates obtained with the different prior assumptions.

The vast majority of our data lies below the $13\,\%\,{\rm mag}^{-1}$ reference level in both maps, but the results differ significantly for low-$E(B-V)$ sightlines. For $E(B-V) \leq 0.1$ mag, the \cite{edenhofer_2024.3d.extinction.map} map leads to several measurements exceeding the reported limits, while for Bayestar19 there is only one outlier at $E(B-V) \approx 0.04$ mag, identified as StarID 14867. In agreement with \cite{panopoulou_2019_p_extinction}, we conclude that the extinction maps limit our ability to constrain polarization efficiencies in diffuse sightlines. Thus, exceeding the reported limits remains a viable scenario in this sample, consistent with previous results \citep{angarita_2023.polarization.efficiency}. 

High-$K$ measurements are shown as diamonds in Fig.~\ref{fig:pol_efficiency}. These measurements span a broad range of polarization efficiencies. Most exhibit polarization efficiencies below $9 \% {\rm mag}^{-1}$ in both extinction maps, although a few measurements approach or exceed the higher reference levels when the \cite{edenhofer_2024.3d.extinction.map} extinction map is used. Consequently, enhanced polarization efficiency, resulting from favorable grain alignment conditions, appears only in individual cases.

\section{Conclusions}
\label{sec:conclusions_future_prospects}

We investigated the challenges of constraining interstellar dust properties from multi-wavelength starlight polarization. By comparing Serkowski fits obtained with uniform and informed priors and examining wavelength-dependent EVPA variations, we assessed the limitations imposed by sparse wavelength coverage, prior assumptions, and the three-dimensional structure of the ISM.

The inferred spectral width, described by $K$, is particularly sensitive to the adopted prior. Uniform priors generally yield larger uncertainties in $K$ than priors favoring the Wilking relation, whereas the uncertainties in $p_{\max}$ and $\lambda_{\max}$ are more comparable. This behavior occurs across all three $\Delta\theta$ groups and is consistent with limited wavelength coverage weakening the constraints on spectral curvature.

Allowing the spectral shape to vary also reveals candidates with polarization profiles narrower than expected from the Wilking relation. For these measurements, uniform-prior fits reproduce the observed polarization more closely than the corresponding informed-prior fits. 

Our illustrative two-cloud models, following \cite{mandarakas_2025}, show that components individually satisfying the Wilking relation can combine to produce a substantially narrower apparent profile over the observed bands. The absence of strong wavelength-dependent EVPA variations does not necessarily exclude LOS superposition. Additionally, the theoretically-derived trends of some grain models seem to be consistent with our observations \citep{cicota_2018.SN_polarization}.

The ratio $p_{\max}/E(B-V)$ depends on the grain properties and magnetic field geometry, and its inferred value is sensitive to the adopted reddening map, particularly for low-extinction sightlines. The high-$K$ objects span a broad range of polarization efficiencies rather than systematically approaching the empirical upper envelopes. This indicates that narrow polarization curves are not a necessary outcome of enhanced grain alignment.

Reliable constraints on dust properties require both improved spectral coverage and independent information about LOS structure. Broad-wavelength spectropolarimetry extending into the near-infrared would better constrain the spectral curvature and wavelength-dependent EVPA without requiring a fixed $K$ or a strict Wilking relation. Spectropolarimeters such as FiberPol-6D \citep{maharana_2024.fiberpol6d} could therefore play an important role in dust studies based on polarization data. Combining these observations with stellar distances from Gaia, three-dimensional dust maps \citep{green_2019.bayestar,edenhofer_2024.3d.extinction.map}, and distance-resolved magnetic-field information \citep{tassis_2018.pasiphae,pelgrims_2024.polarization.tomography} will help distinguish intrinsic grain-property variations from line-of-sight integration effects.

\begin{acknowledgements}
I sincerely thank the anonymous reviewer for their constructive comments and suggestions, which were invaluable in the revision of this manuscript. My deepest gratitude to K. Tassis, P. F. Goldsmith, G. V. Panopoulou, I. Liodakis, D. Blinov, S. Kiehlmann, and A. Synani for providing comments on the draft. I am grateful to N. Mandarakas, and B. Hensley for fruitful discussions. Many thanks to T. Dharmawardena, and A. Saydjari for several discussions on 3D dust extinction maps during the Hubble Symposium 2024, which took place at Caltech. I am also grateful to Dr. Christidis for inspiring discussions. This work was supported by NSF grant AST-2109127. Support for this work was provided by NASA through the NASA Hubble Fellowship grant \#~HST-HF2-51566.001 awarded by the Space Telescope Science Institute, which is operated by the Association of Universities for Research in Astronomy, Inc., for NASA, under contract NAS5-26555. In this work, we have made use of the following software: BayesianTools \citep{BayesianTools}, data.table \citep{R.data.table}, ggplot2 \citep{R.ggplot2}.	
\end{acknowledgements}

\bibliographystyle{aa}
\bibliography{bibliography}

%---------------------------------------APPENDIX-------------------------
\begin{appendix}

\section{Polarization likelihoods}
\label{appendix:likelihoods}

\begin{figure}
	\centering
	\includegraphics[width=\hsize]{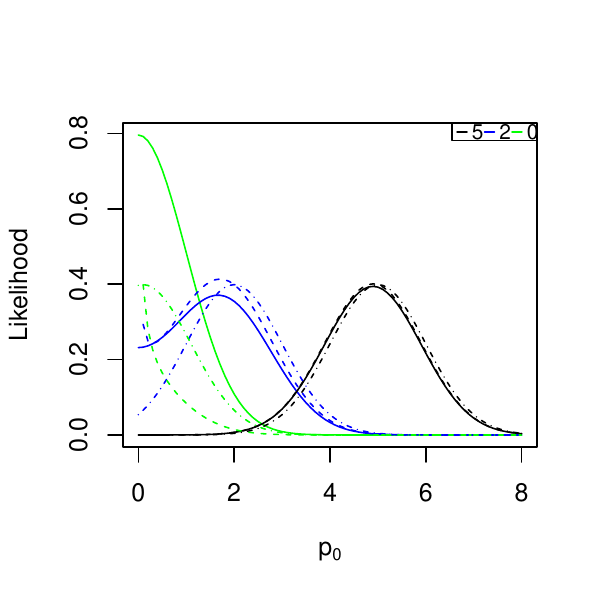}
	\caption{Normalized likelihood functions of the intrinsic (true or unbiased) polarization fraction for different S/N.   Green curves correspond to S/N=0.1, while blue and black curves to S/N = 2, and 5 respectively, as shown in the legend. Solid, and dashed curves correspond to the total (Eq.\ref{eq:likelihood}) and asymptotic (Eq.~\ref{eq:likelihood_asymptotic}) polarization likelihoods, while dotted-dashed curves correspond to Gaussians. When S/N = 5, the asymptotic likelihood expression is nearly identical to the total likelihood. }
	\label{fig:likelihoods}
\end{figure}

We demonstrate the properties of the polarization likelihoods at different S/N. We set the observational uncertainty equal to $\sigma = 1 \%$ and assumed different observed $p$ values. In this way, we control the S/N of the measurements as follows: for $p=0.1$ we obtain that S/N=0.1, while for $p=2$, we get that S/N=2. 

Fig.~\ref{fig:likelihoods} shows different normalized likelihood functions for S/N equal to 0.1 (green), 2 (blue), and 5 (black). Solid curves correspond to the total polarization likelihood that is a Rice distribution (Eq.~\ref{eq:likelihood}, and  Fig. 2 in \citealt{vaillancourt_2006}). Dashed curves show the asymptotic expression of the polarization likelihood (Eq.~\ref{eq:likelihood_asymptotic}), while dashed-dotted curves Gaussian profiles. For S/N=5, the different likelihoods are very close to each other, implying that the polarization likelihood approximates a Gaussian. However, the asymptotic likelihood has a slightly different amplitude than a Gaussian, which is also evident in Fig.~\ref{fig:likelihoods} where the peak of the Gaussian (dashed-dotted black curve) is slightly shifted compared to the other two, which are nearly identical. 

For S/N < 3 (green and blue curves), both the Gaussian and the asymptotic likelihoods deviate from the polarization likelihood significantly. Overall, we found that the asymptotic form is accurate for S/N $\geq 4$, while a Gaussian for S/N $\geq 5$. We conclude that the asymptotic likelihood shown in Eq.~(\ref{eq:likelihood_asymptotic}) more accurately approximates the Rice distribution than a Gaussian at high S/N. 

\section{High-$K$ measurements}
\label{appendix:high_k_stars}

\begin{table*}
\caption{Selected high-$K$ candidates and their uniform-prior Serkowski parameters.}
\label{tab:highK_measurements}
\centering
\scriptsize
\setlength{\tabcolsep}{2pt}
\renewcommand{\arraystretch}{1.18}
\resizebox{\textwidth}{!}{%
\begin{tabular}{@{}rrrllccccccrrr@{}}
\hline\hline
starID & $l$ & $b$ & Field & Type & Inst. & Cut & Ref. & $p_{\max}$ & $\lambda_{\max}$ & $K$ & $\Delta\theta$ & $\chi^2_{p,\mathrm{U}}$ & $\chi^2_{p,\mathrm{I}}$ \\
 & (deg) & (deg) & & & & & & (\%) & ($\mu$m) & & & & \\
\hline
22480 & 118.216 & $+4.968$ & Berkeley 59 & OC & 1 & $3\sigma$ & (a) & $4.86_{-0.14}^{+0.14}$ & $0.65_{-0.02}^{+0.01}$ & $4.61_{-1.41}^{+1.31}$ & $1.22$ & $0.11$ & $11.00$ \\
22484 & 118.267 & $+4.987$ & Berkeley 59 & OC & 1 & $3\sigma$ & (a) & $4.29_{-0.19}^{+0.18}$ & $0.67_{-0.01}^{+0.01}$ & $7.82_{-2.14}^{+1.47}$ & $1.79$ & $0.44$ & $12.89$ \\
22500 & 118.471 & $+5.003$ & Berkeley 59 & OC & 1 & $3\sigma$ & (a) & $8.42_{-0.26}^{+0.25}$ & $0.62_{-0.01}^{+0.01}$ & $6.94_{-1.41}^{+1.28}$ & $2.98$ & $0.06$ & $26.83$ \\
22698 & 190.415 & $-0.355$ & LDN 1570 & DC & 1 & $3\sigma$ & (b) & $3.98_{-0.11}^{+0.11}$ & $0.62_{-0.02}^{+0.01}$ & $4.67_{-1.52}^{+1.22}$ & $0.00$ & $0.21$ & $16.00$ \\
22697 & 190.479 & $-0.391$ & LDN 1570 & DC & 1 & $2\sigma$ & (b) & $4.61_{-0.34}^{+0.37}$ & $0.60_{-0.05}^{+0.03}$ & $6.26_{-3.02}^{+2.41}$ & $1.64$ & $0.38$ & $6.63$ \\
24161 & 6.934 & $-2.203$ & Sh 2-29 & HII & 2 & $2\sigma$ & (c) & $4.31_{-0.34}^{+0.43}$ & $0.61_{-0.07}^{+0.03}$ & $6.11_{-3.31}^{+2.71}$ & $1.24$ & $0.89$ & $5.84$ \\
44467 & 348.195 & $-1.057$ & RCW 121 & HII & 2 & $2\sigma$ & (d) & $3.33_{-0.12}^{+0.13}$ & $0.61_{-0.06}^{+0.03}$ & $4.27_{-2.07}^{+2.02}$ & $3.38$ & $0.42$ & $6.11$ \\
44468 & 348.230 & $-1.033$ & RCW 121 & HII & 2 & $2\sigma$ & (d) & $5.36_{-0.13}^{+0.14}$ & $0.65_{-0.04}^{+0.02}$ & $5.60_{-2.52}^{+2.42}$ & $1.53$ & $0.34$ & $7.01$ \\
44475 & 348.266 & $-0.956$ & RCW 121 & HII & 2 & $3\sigma$ & (d) & $2.84_{-0.13}^{+0.10}$ & $0.63_{-0.01}^{+0.01}$ & $7.94_{-2.85}^{+1.39}$ & $1.51$ & $0.59$ & $20.24$ \\
44505 & 348.246 & $-1.023$ & RCW 121 & HII & 2 & $3\sigma$ & (d) & $2.90_{-0.18}^{+0.23}$ & $0.75_{-0.02}^{+0.05}$ & $7.59_{-2.75}^{+1.65}$ & $3.63$ & $0.97$ & $11.65$ \\
44533 & 348.231 & $-1.021$ & RCW 121 & HII & 2 & $2\sigma$ & (d) & $1.93_{-0.10}^{+0.11}$ & $0.66_{-0.03}^{+0.03}$ & $6.76_{-3.57}^{+2.26}$ & $0.98$ & $0.88$ & $7.46$ \\
44619 & 348.238 & $-1.012$ & RCW 121 & HII & 2 & $2\sigma$ & (d) & $3.76_{-0.15}^{+0.15}$ & $0.61_{-0.04}^{+0.02}$ & $4.68_{-2.17}^{+2.17}$ & $2.92$ & $0.24$ & $6.02$ \\
44662 & 348.216 & $-1.032$ & RCW 121 & HII & 2 & $2\sigma$ & (d) & $3.93_{-0.22}^{+0.54}$ & $0.58_{-0.07}^{+0.03}$ & $7.01_{-3.04}^{+2.08}$ & $0.98$ & $0.88$ & $18.13$ \\
44812 & 348.238 & $-1.036$ & RCW 121 & HII & 2 & $2\sigma$ & (d) & $3.77_{-0.10}^{+0.10}$ & $0.62_{-0.03}^{+0.02}$ & $4.15_{-2.03}^{+1.91}$ & $1.93$ & $0.21$ & $6.07$ \\
44846 & 348.163 & $-1.027$ & RCW 121 & HII & 2 & $2\sigma$ & (d) & $2.27_{-0.09}^{+0.09}$ & $0.62_{-0.02}^{+0.02}$ & $6.80_{-2.87}^{+2.02}$ & $5.40$ & $0.31$ & $10.11$ \\
44854 & 348.176 & $-0.953$ & RCW 121 & HII & 2 & $2\sigma$ & (d) & $1.49_{-0.09}^{+0.09}$ & $0.64_{-0.03}^{+0.02}$ & $6.94_{-3.34}^{+2.14}$ & $2.16$ & $0.67$ & $7.78$ \\
48510 & 6.084 & $-1.308$ & NGC 6530 & OC & 3 & $2\sigma$ & (e) & $1.00_{-0.09}^{+0.08}$ & $0.57_{-0.02}^{+0.02}$ & $5.72_{-2.22}^{+1.92}$ & $2.34$ & $0.12$ & $10.31$ \\
48568 & 186.065 & $-13.190$ & NGC 1817 & OC & 1 & $2\sigma$ & (f) & $0.93_{-0.12}^{+0.10}$ & $0.60_{-0.02}^{+0.03}$ & $6.47_{-3.42}^{+2.12}$ & $3.24$ & $0.41$ & $10.18$ \\
48581 & 186.202 & $-13.242$ & NGC 1817 & OC & 1 & $3\sigma$ & (f) & $1.23_{-0.12}^{+0.10}$ & $0.61_{-0.01}^{+0.02}$ & $7.12_{-2.59}^{+1.73}$ & $2.35$ & $0.26$ & $16.01$ \\
48600 & 186.216 & $-13.020$ & NGC 1817 & OC & 1 & $2\sigma$ & (f) & $1.14_{-0.12}^{+0.11}$ & $0.57_{-0.02}^{+0.02}$ & $5.20_{-2.64}^{+1.88}$ & $2.83$ & $0.64$ & $10.28$ \\
48664 & 123.085 & $-13.203$ & Cas. Alessi 1 & OC & 1 & $2\sigma$ & (g) & $1.04_{-0.12}^{+0.10}$ & $0.59_{-0.02}^{+0.02}$ & $7.46_{-3.43}^{+1.74}$ & $0.83$ & $0.47$ & $13.30$ \\
49072 & 104.013 & $+6.374$ & NGC 7160 & OC & 3 & $2\sigma$ & (h) & $0.92_{-0.09}^{+0.08}$ & $0.53_{-0.02}^{+0.02}$ & $6.91_{-2.95}^{+1.95}$ & $2.08$ & $0.51$ & $12.79$ \\
49191 & 226.514 & $-2.321$ & NGC 2345 & OC & 1 & $3\sigma$ & (i) & $2.74_{-0.16}^{+0.15}$ & $0.59_{-0.01}^{+0.02}$ & $7.31_{-1.99}^{+1.58}$ & $14.71$ & $0.11$ & $23.02$ \\
49225 & 226.487 & $-2.403$ & NGC 2345 & OC & 1 & $3\sigma$ & (i) & $2.85_{-0.12}^{+0.12}$ & $0.61_{-0.01}^{+0.01}$ & $5.43_{-0.87}^{+0.82}$ & $5.81$ & $0.84$ & $31.09$ \\
49288 & 227.172 & $-1.517$ & NGC 2345 & OC & 1 & $3\sigma$ & (i) & $2.06_{-0.12}^{+0.11}$ & $0.62_{-0.01}^{+0.01}$ & $4.41_{-1.22}^{+1.13}$ & $7.36$ & $0.50$ & $12.27$ \\
49321 & 226.980 & $-1.441$ & NGC 2345 & OC & 1 & $2\sigma$ & (i) & $1.60_{-0.11}^{+0.11}$ & $0.62_{-0.03}^{+0.03}$ & $5.36_{-2.03}^{+2.05}$ & $2.28$ & $0.53$ & $8.27$ \\
49325 & 226.932 & $-1.427$ & NGC 2345 & OC & 1 & $2\sigma$ & (i) & $2.73_{-0.20}^{+0.19}$ & $0.63_{-0.05}^{+0.02}$ & $6.11_{-3.41}^{+2.59}$ & $9.28$ & $0.81$ & $6.48$ \\
49341 & 226.963 & $-1.516$ & NGC 2345 & OC & 1 & $2\sigma$ & (i) & $1.16_{-0.18}^{+0.16}$ & $0.53_{-0.05}^{+0.03}$ & $5.76_{-3.01}^{+2.44}$ & $12.01$ & $0.60$ & $8.75$ \\
49344 & 226.967 & $-1.538$ & NGC 2345 & OC & 1 & $3\sigma$ & (i) & $2.08_{-0.17}^{+0.16}$ & $0.53_{-0.01}^{+0.01}$ & $5.80_{-1.70}^{+1.41}$ & $10.60$ & $0.06$ & $19.99$ \\
49347 & 226.974 & $-1.510$ & NGC 2345 & OC & 1 & $2\sigma$ & (i) & $1.65_{-0.13}^{+0.13}$ & $0.55_{-0.04}^{+0.03}$ & $3.96_{-1.73}^{+1.80}$ & $1.56$ & $0.06$ & $5.66$ \\
\hline
\end{tabular}%
}
\tablefoot{The Serkowski parameters are the uniform-prior posterior medians. U and I denote uniform and informed priors. The Cut column denotes the relative-precision selection: $\operatorname{med}(x)\geq n\,U_{68}(x)/2$ for both $x=K$ and $\lambda_{\max}$, with $n=2$ or 3 and $U_{68}=x_{84}-x_{16}$. Field codes: OC = open cluster; DC = dark cloud; HII = H\,\textsc{ii} region. Instrument codes: 1 = AIMPOL \citep{rautela_2004.aimpol}; 2 = IAGPOL \citep{magalhaes_1996.iagpol}; 3 = VMI \citep{topasna_2013.vmi.polarimeter}.}
\tablebib{
(a) \cite{eswaraiah_2012.polarimetry.young.open.cluster};
(b) \cite{eswaraiah_2013.ldn1570.polarimetry};
(c) \cite{santos_2014.sh2.29.polarization};
(d) \cite{seronnavarrete_2016.rcw121.polarimetry};
(e) \cite{topasna_2020.ngc6530.extinction};
(f) \cite{singh_2020.ngc1817.polarization} ;
(g) \cite{singh_2020.casado.alessi1.polarization};
(h) \cite{topasna_2021.ngc7160.extinction.polarization};
(i) \cite{singh_2022.ngc2345.grain.alignment}.}
\end{table*}

\begin{figure*}
	\centering
	\includegraphics[width=0.8\hsize]{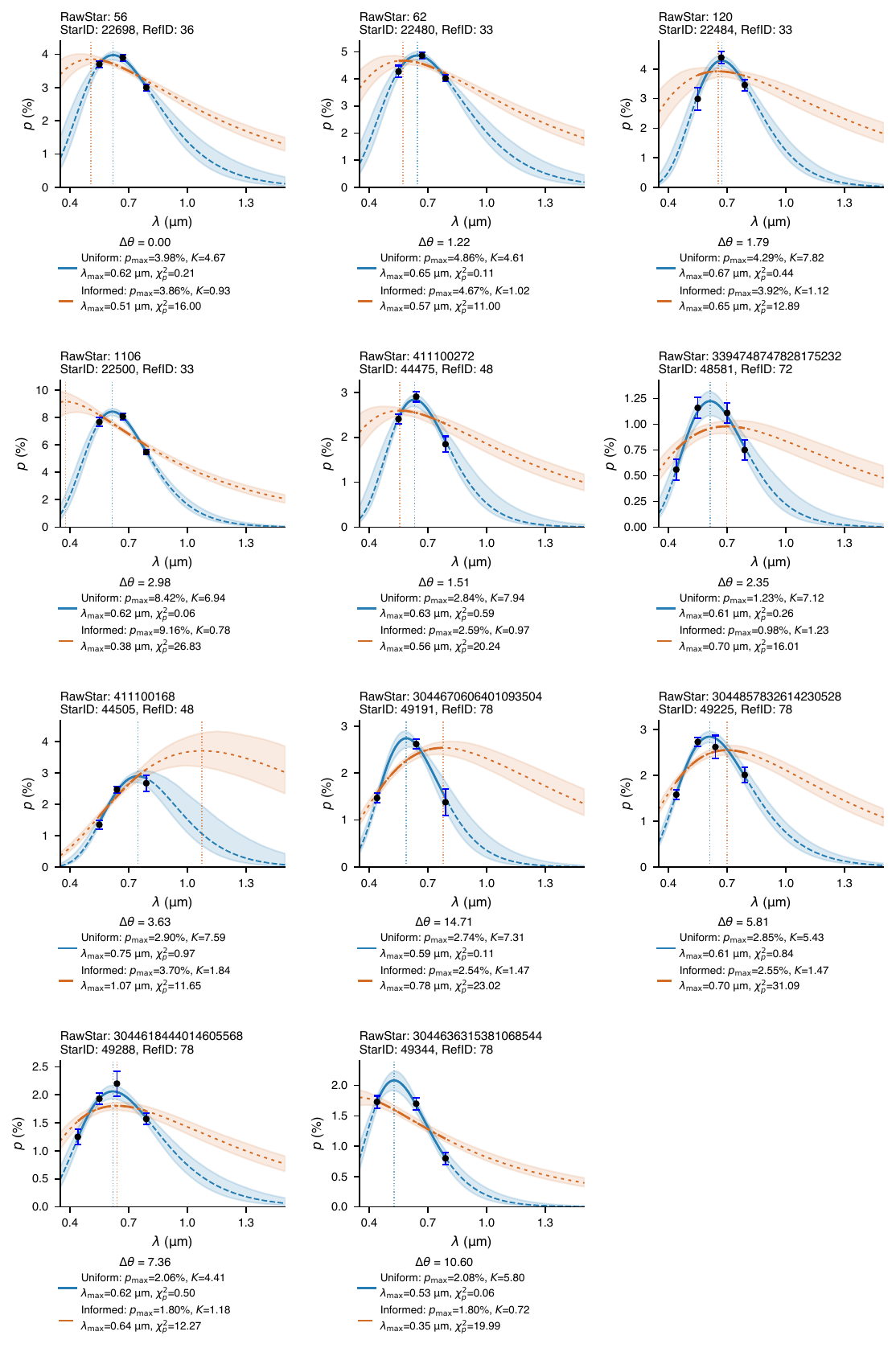}
	\caption{Polarization spectra of the 11 high-$K$ measurements. Symbols and line styles are the same as in Fig.~\ref{fig:serkowski_fits_examples}. Panel titles identify each measurement by its original catalogue identifiers. Legends list the Serkowski parameters, $\chi^2_p$ for both priors, and $\Delta \theta$.}
	\label{fig:extreme_k_spectra}
\end{figure*}

Table \ref{tab:highK_measurements} lists the 30 measurements with elevated K values selected in Sect. \ref{sec:wilking_relation_deviations}. The table lists the unique identifier of each star, its Galactic coordinates, the field in which the object is located (or to which it is in close proximity), the field type, the instrument used for the measurements, the statistical precision with which $K$ and \lmax\ are constrained (as described in Sect.~\ref{sec:wilking_relation_deviations}), the obtained Serkowski parameters, and the corresponding references. Fig.~\ref{fig:extreme_k_spectra} shows the polarization spectra of the 11 candidates that satisfy the $3\sigma$ selection criterion, including the fits obtained with both uniform and informed priors.

\end{appendix}

\end{document}